\documentclass[aps,prl,twocolumn,groupedaddress]{revtex4}
\usepackage{graphicx}
\usepackage{amsmath}
\usepackage{dcolumn}
\usepackage{bm}
\usepackage{times}

\begin{document}
\title{Quantum reactive scattering of ultracold NH($X\,^3\Sigma^-$) radicals in a magnetic trap}

\author{Liesbeth M.~C.~Janssen}
\email[Electronic mail: ]{ljanssen@science.ru.nl}
\affiliation{Radboud University Nijmegen, Institute for Molecules and Materials,
Heyendaalseweg 135, 6525 AJ Nijmegen, The Netherlands}
\author{Ad van der Avoird}
\affiliation{Radboud University Nijmegen, Institute for Molecules and Materials,
Heyendaalseweg 135, 6525 AJ Nijmegen, The Netherlands}
\author{Gerrit C.~Groenenboom}
\email[Electronic mail: ]{gerritg@theochem.ru.nl}
\affiliation{Radboud University Nijmegen, Institute for Molecules and Materials,
Heyendaalseweg 135, 6525 AJ Nijmegen, The Netherlands}

\date{\today}

\begin{abstract}
We investigate the ultracold reaction dynamics of magnetically trapped
NH($X\,^3\Sigma^-$) radicals using rigorous quantum scattering calculations
involving three coupled potential energy surfaces. We find that the reactive NH + NH 
cross section is driven by a short-ranged collisional mechanism, and its magnitude
is only weakly dependent on magnetic field strength. Unlike most ultracold
reactions observed so far, the NH + NH scattering dynamics is non-universal.
Our results indicate that chemical
reactions can cause more trap loss than spin-inelastic NH + NH collisions,
making molecular evaporative cooling more difficult than previously
anticipated.
\end{abstract}

\maketitle

The ability to produce and trap molecules at sub-kelvin temperatures offers
numerous exciting possibilities in chemistry.  Recent experiments have
demonstrated that ultracold chemical reactions can be efficiently manipulated
using an external electromagnetic field \cite{ni:10,knoop:10,miranda:11},
providing new tools to control reaction pathways and rate constants. The
pronounced quantum behavior of ultracold matter may also lead to novel
phenomena such as ``superchemistry", a process in which an atomic and molecular
Bose-Einstein condensate (BEC) are coherently coupled to stimulate chemical
reactivity \cite{heinzen:00}. Up to the present, however, studies on ultracold
chemical reactions have focused only on (bi-)alkali-metal systems with a rather
limited chemistry. Moreover, most of the observed ultracold reactive processes
exhibit universal behavior, i.e., the dynamics are completely determined by
long-range interactions \cite{idziaszek:10}. Cold reactive collisions in the
non-universal regime, as well as cold reactions involving non-alkali and
open-shell molecules, are still largely unexplored.

Cold chemistry is also relevant in the context of evaporative and sympathetic
cooling. These second-stage cooling methods, which rely on strong elastic
collisions between trapped particles, represent the final step towards full
quantum degeneracy and Bose-Einstein condensation \cite{ketterle:96}.  As a
rule of thumb, the ratio between elastic and non-elastic (inelastic and
reactive) cross sections should be at least two orders of magnitude in order
for second-stage cooling to succeed.  For polar molecules, whose properties are
expected to find wide applications in ultracold physics
\cite{demille:02,micheli:06,zelevinsky:08,carr:09,chin:09}, cooling into the
quantum-degenerate regime is yet to be achieved experimentally. Various
theoretical studies suggest that molecular second-stage cooling is feasible
\cite{wallis:09,gonzalez:11,wallis:11,hummon:11,
zuchowski:11,janssen:11a,tscherbul:11,suleimanov:12}, but these investigations
are based on the assumption that chemical reactions are strongly suppressed in
a (magnetically trapped) spin-polarized gas \cite{krems:08} and that the
dynamics evolves on a single non-reactive potential. To our knowledge, this
assumption has not yet been validated by explicit reactive scattering
calculations. 

In this Letter, we present rigorous coupled-channels calculations for the
reactive NH($X\,^3\Sigma^-$) +  NH($X\,^3\Sigma^-$) system in the presence of a
magnetic field. Ultracold reactive NH + NH collisions can yield as many as
eight different product arrangements \cite{lai:03,poveda:09}, making it a
versatile system for cold chemistry studies. NH has already been cooled to
millikelvin temperatures and stored in a magnetic trap using Stark deceleration
\cite{hoekstra:07,riedel:11} and buffer gas cooling techniques
\cite{campbell:07,hummon:08,hummon:11}, and earlier theoretical studies --
based on non-reactive scattering calculations -- have indicated that NH is a
promising candidate for second-stage cooling experiments
\cite{wallis:09,gonzalez:11,wallis:11,hummon:11,
zuchowski:11,janssen:11,janssen:11a,suleimanov:12}.  Within a magnetic trap, NH
is polarized in the low-field-seeking state $|S_{\rm{NH}}=1,
M_{S_{\rm{NH}}}=1\rangle$, with $S_{\rm{NH}}$ denoting the total electronic
spin and $M_{S_{\rm{NH}}}$ its projection onto the magnetic field axis. A
collision complex of two such molecules is in the high-spin quintet
$|S=2,M_S=2\rangle$ state, with $S$ and $M_S$ referring to the dimer spin
quantum numbers. Spin-inelastic NH + NH collisions can change either the $M_S$
quantum number of the quintet state or the total spin $S$ to produce singlet
($S=0$) or triplet ($S=1$) complexes.  Since the $S=0$ and 1 states are
chemically reactive \cite{lai:03}, $S$-changing transitions may also initiate
chemical reaction.

In order to study the reaction dynamics of cold NH radicals, we have employed a
single-arrangement quantum reactive scattering method that allows for the
calculation of total NH + NH reaction probabilities.  The cross sections are
obtained from full quantum scattering calculations on three ($S=0,1,2$) coupled
potential energy surfaces.  Our results indicate that chemical reactions can
cause more trap loss than inelastic collisions, implying that reactive channels
must be taken into account when assessing the feasibility of evaporative and
sympathetic cooling. We also find that the total reaction probability is
strongly dependent on the details of the (short-range) interaction potentials,
providing one of the first examples of non-universal cold chemistry.

We focus on collisions between two magnetically trapped bosonic
$^{15}$NH($X\,^3\Sigma^-$) molecules and treat the monomers as rigid rotors.
The coordinate system consists of the intermolecular vector $\bm{R}$, with
length $R$, the polar angles $\theta_A$ and $\theta_B$ of the monomers, and the
dihedral angle $\phi$. The scattering Hamiltonian is given by
\begin{eqnarray}
\label{eq:H}
\hat{H} &=& -\frac{\hbar^2}{2\mu R} \frac{\partial^2}{\partial R^2}R +
            \frac{\hat{L}^2}{2\mu R^2}  \nonumber \\
       & & +
            \sum_{S,M_S} |S,M_S\rangle V_S(R,\theta_A,\theta_B, \phi) \langle S,M_S| 
            \nonumber \\
       & & + 
             V_{\rm{magn.dip}}(\bm{R},\hat{\bm{S}}_A,\hat{\bm{S}}_B) 
           + \hat{H}_A + \hat{H}_B,
\end{eqnarray}
where $\mu$ is the reduced mass of the complex, $\hat{L}^2$ is the angular
momentum operator associated with rotation of $\bm{R}$,
$V_S(R,\theta_A,\theta_B,\phi)$ is the potential energy surface for total spin
$S$, $V_{\rm{magn.dip}}(\bm{R},\hat{\bm{S}}_A,\hat{\bm{S}}_B)$ is the
intermolecular magnetic dipole interaction between the two monomer triplet
spins, and $\hat{H}_A$ and $\hat{H}_B$ are the Hamiltonians of the individual
monomers. The latter account for the monomer rotation, intramolecular spin-spin
coupling, spin-rotation coupling, and Zeeman interaction \cite{janssen:11a}.
The terms that couple the $S=0, 1$, and 2 potentials and that ultimately drive
spin-changing processes are the intermolecular magnetic dipole interaction and
the intramolecular spin-spin and spin-rotation couplings.

\begin{figure}
\centering
\includegraphics[width=8cm]{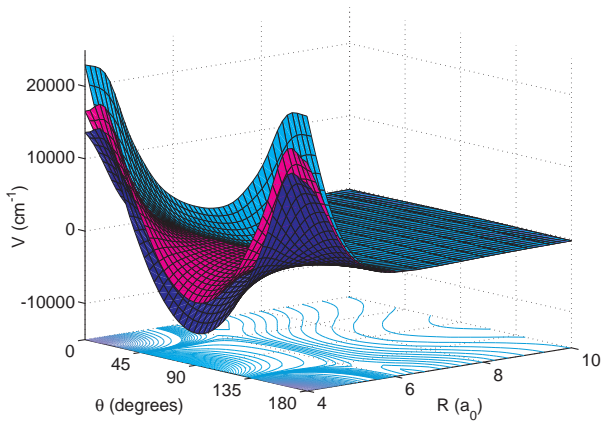}
\caption
  {\label{fig:PESs}
Cuts through the quintet (cyan), triplet (magenta), and singlet (dark blue)
potential energy surfaces of NH($X\,^3\Sigma^-$) -- NH($X\,^3\Sigma^-$)
for $\phi = 180^{\circ}$ and $\theta_A=\theta_B=\theta$.
}
\end{figure}

The NH--NH potential energy surfaces have been obtained from high-level
\textit{ab initio} calculations, as described previously \cite{janssen:09}.
Figure \ref{fig:PESs} shows a cut through the three NH--NH potentials for $\phi
= 180^{\circ}$ and $\theta_A=\theta_B=\theta$.  It can be seen that the singlet
and triplet potentials are strongly attractive at small intermolecular
distances, which is due to their chemically reactive nature.  In view of these
deep potential energy wells we may assume that, once a reactive singlet or triplet
NH--NH complex is formed, the system readily undergoes exoergic chemical
rearrangement.  For instance, NH + NH may react into N$_2$H$_2$  (provided that
a third body can dissipate the excess kinetic energy) or into a binary product
configuration such as N$_2$ + H$_2$ \cite{poveda:09}.  In order to calculate
the total reaction probability, we consider only the NH + NH reactant
arrangement and apply ``capture" boundary conditions at short range.
That is, at a sufficiently small value of
the radial coordinate $R$, we allow flux to disappear into reactive channels.
Such an approach is commonly used in (reactive) scattering problems involving
deep potential energy wells \cite{klippenstein:08}. 
We note that not all collisions on the singlet and triplet potentials are reactive, 
since these surfaces are repulsive for certain geometries. Hence, $S$-changing
collisions will not necessarily lead to chemical reaction.
This is also contained in our boundary
conditions (see Supplementary Material).
Collisions occurring on the quintet potential are entirely non-reactive.

We have developed a novel reactive scattering algorithm based on the
renormalized Numerov propagator to calculate the relevant collision cross
sections. Our algorithm amounts to solving the coupled-channels equations for
\textit{two} sets of solutions, and subsequently applying reactive scattering
boundary conditions to obtain the $S$-matrix. Details can be found in the
Supplementary Material.  We used a symmetry-adapted channel basis set with even
permutation symmetry and even parity, and a space-fixed total angular momentum
projection of $\mathcal{M}=2$ (see also Supplementary Material). This basis
allows for $s$-wave collisions between identical bosonic molecules in the same
initial quantum state. 
The radial grid ranged from $4.5$ to $1500$ $a_0$, with
a minimum of 10 grid points per smallest (local) de Broglie wavelength.  The
cross sections were calculated for collision energies $E$ of 10$^{-6}$ to 1 K
and magnetic field strengths $B$ of $10^{-1}$ to $10^{4}$ G.

\begin{figure}[t!]
\centering
\includegraphics[width=8cm]{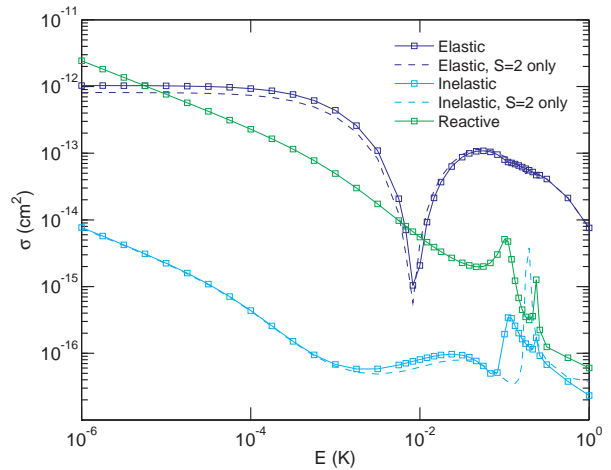}
\caption
  {\label{fig6:res1}
Cross sections for magnetically trapped $^{15}$NH + $^{15}$NH as a function of
collision energy, calculated for a magnetic field strength of 1 G. The solid
lines were obtained by including all three potentials in the scattering
calculations, while the dashed lines were obtained using only the non-reactive
quintet potential.
}
\end{figure}

Figure \ref{fig6:res1} shows the elastic, spin-inelastic, and reactive cross
sections for two magnetically trapped NH molecules as a function of the
collision energy for $B=1$ G.  We find that the elastic cross section at small
$E$ is constant as a function of energy, consistent with Wigner's threshold law
for $s$-wave elastic scattering \cite{wigner:48}, $\sigma_{\rm{elastic}}
\propto E^{L_{\rm{in}} + L_{\rm{out}}}$. Here $L_{\rm{in}}$ and $L_{\rm{out}}$
denote the partial waves for the incoming and outgoing channels, respectively.
The inelastic and reactive cross sections exhibit $E^{-1/2}$ threshold
behavior, as expected from the $E^{L_{\rm{in}}-1/2}$ law for exoergic $s$-wave
($L_{\rm{in}}=0$) collisions \cite{wigner:48}.  In order to compare our results
with the non-reactive case, we have also plotted the cross sections obtained
from scattering calculations on the non-reactive quintet potential.  As can be
seen in Fig.\ \ref{fig6:res1}, the inclusion of chemically reactive ($S=0$ and
$1$) potentials has an almost negligible effect on the elastic and
spin-inelastic cross sections, confirming our earlier expectations reported in
Ref.\ \cite{janssen:11}. We thus conclude that most of the non-reactive,
inelastic trap loss occurs on the quintet surface, and that collisions on the
$S=0$ and 1 potentials are almost 100\% reactive.

Figure \ref{fig6:res1} also indicates that (for $B=1$ G) the
elastic-to-reactive cross section ratio is much smaller than the
elastic-to-inelastic ratio, the difference being 2 to 3 orders of magnitude.
This leads us to reconsider the prospects for molecular evaporative cooling.
While previous studies based on non-reactive scattering calculations have found
that evaporative cooling of NH is feasible
\cite{janssen:11,janssen:11a,suleimanov:12}, our present results indicate that
chemical reactions can pose a major constraint on the efficiency of the cooling
process. Thus, second-stage cooling of NH might be more difficult than
previously expected.  We will return to this topic later, when discussing the
effect of uncertainties in the potentials.

\begin{figure}[t!]
\centering
\includegraphics[width=8cm]{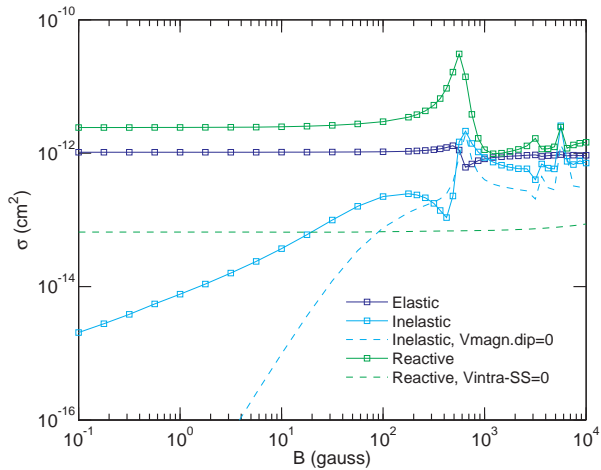}
\caption
 {\label{fig6:res2}
Cross sections for magnetically trapped $^{15}$NH + $^{15}$NH as a function of
magnetic field strength, calculated at $E = 10^{-6}$ K. The spin-inelastic
cross sections obtained with the intermolecular magnetic dipole term switched
off, and the reactive cross sections obtained with the intramolecular spin-spin
coupling switched off, are also plotted.
}
\end{figure}

In Fig.\ \ref{fig6:res2}, we present the collision cross sections as a function
of magnetic field for a collision energy of 10$^{-6}$ K.  We have already
established in previous work \cite{janssen:11a, janssen:11b} that the
intermolecular magnetic dipolar term is the dominant source of inelastic trap
loss at ultralow energies. Indeed, when discarding the intermolecular magnetic
dipole-dipole coupling ($V_{\rm{magn.dip}}$) in our reactive scattering
calculations, the spin-inelastic cross section decreases by several orders of
magnitude.  For the \textit{reactive} cross sections, however, we find that the
main contribution comes from the \textit{intramolecular} spin-spin coupling
(denoted as $V_{\rm{intra-SS}}$ in Fig.\ \ref{fig6:res2}).  The effect of the
intermolecular term $V_{\rm{magn.dip}}$ on the chemical reactivity is almost
negligible.  This can also be understood by considering that the intermolecular
spin-spin interaction is long-ranged \cite{janssen:11b}, while a chemical
reaction can only proceed when the reactants approach each other to a very
short distance. Hence, the intramolecular spin-spin coupling, which acts
through the potential anisotropy at short range \cite{krems:04b,krems:03a},
plays the most important role in enabling the reaction to occur.

It may also be seen that the reactive cross section at small fields ($B < 100$
G) is constant as a function of $B$. This is essentially a consequence of the
$E^{L_{\rm{in}}-1/2}$ law for exoergic collisions: the threshold behavior is
determined only by the centrifugal barrier in the entrance channel, and the
energy of the outgoing (reactive) channel has no effect on the cross section.
Note that a different threshold law applies for short-range-induced
\textit{inelastic} processes, for which the centrifugal barrier and energy of
the outgoing channel do play a role \cite{volpi:02}.

\begin{figure}[t]
\centering
\includegraphics[width=8cm]{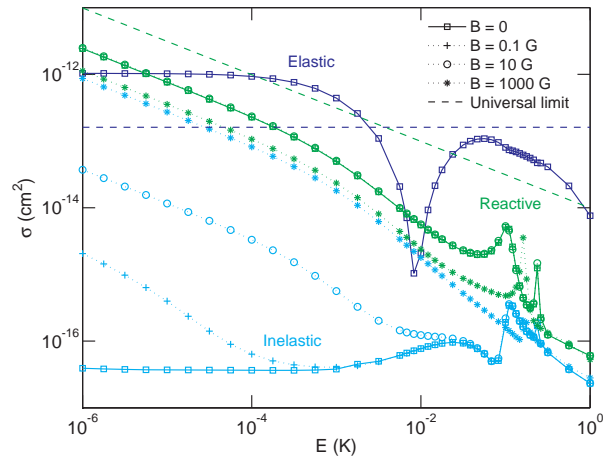}
\caption
  {\label{fig6:res3}
Cross sections for magnetically trapped $^{15}$NH + $^{15}$NH as a function of
collision energy, calculated for several magnetic field strengths. The elastic
cross sections are the same for all magnetic fields considered, and the
reactive cross sections are the same for $B=0, 0.1$, and 10 G.  The results
of the universal quantum-defect model are also shown.
}
\end{figure}

Figure \ref{fig6:res3} compares the scattering results for different magnetic
field strengths, including $B=0$, as a function of collision energy.  We find
that the reactive cross sections are independent of magnetic field for $B <
100$ G at virtually all energies considered, and the \{$B^0$,
$E^{L_{\rm{in}}-1/2}$\} regime extends down to $B=0$.  The \textit{inelastic}
cross sections, however, are different for all $B$ values, and they become
constant as a function of energy for $B=0$ 
\footnote{The threshold behavior for the $V_{\rm{magn.dip}}$-induced inelastic
cross section in a nonzero magnetic field has already been discussed in Ref.\
\cite{janssen:11b}.  There it was shown that the cross section for $B > 0$
shows an $E^0$ regime for moderately low collision energies, and an $E^{-1/2}$
regime for ultralow energies.  As the magnetic field decreases, the $E^0$
regime extends to an increasingly large range of $E$.  Using a similar approach
based on the Born approximation, it can be shown that the inelastic cross
section in \textit{zero} field is constant for \textit{all} energies below some
critical $E$. That is, the $E^0$ regime extends to $E=0$ for $B=0$ (to be
published).}.
Thus, the reactive and inelastic
cross sections show fundamentally different behavior in the ultracold regime.

Let us now consider ultracold NH + NH scattering in the universal limit.
Knowledge of the degree of universality is useful to understand and interpret
the (generally complex) full-dimensional reaction dynamics in terms of simple
few-parameter models.  In order to establish whether the ultracold chemistry
of NH + NH is universal, we compare our scattering results with the
single-channel quantum-defect model of Idziaszek and Julienne
\cite{idziaszek:10}. Details of this model are given in the Supplementary
Material. In the universal limit, all scattering flux that reaches the short
range disappears into reactive channels.  Figure \ref{fig6:res3} compares the
results of the universal quantum-defect model with those of the numerical
coupled-channels calculations.  In the $s$-wave regime ($E < 10^{-3}$ K), the
elastic cross sections are underestimated by the universal model, while the
reactive cross sections are clearly overestimated. The \textit{ratio} between
the universal elastic and reactive cross sections differs by more than one
order of magnitude (a factor of $\approx26$)  from the numerical data.  These
differences suggest that the NH + NH reaction dynamics is non-universal, because
a significant fraction of the incident flux is reflected at short range.
It should be noted, however, that the quantum-defect results apply to
single-channel scattering on a single isotropic potential, while the numerical
data have been obtained from multi-channel calculations on three coupled
anisotropic potential energy surfaces.  Nevertheless, the collision dynamics in
the ultracold regime is $s$-wave dominated, and the effective NH--NH long-range
potential -- which is the same for all three spin states -- is governed mainly
by the isotropic interaction.  Thus, the single-channel quantum-defect model
should be applicable to NH--NH in the universal regime.

\begin{figure}
\centering
\includegraphics[width=8cm]{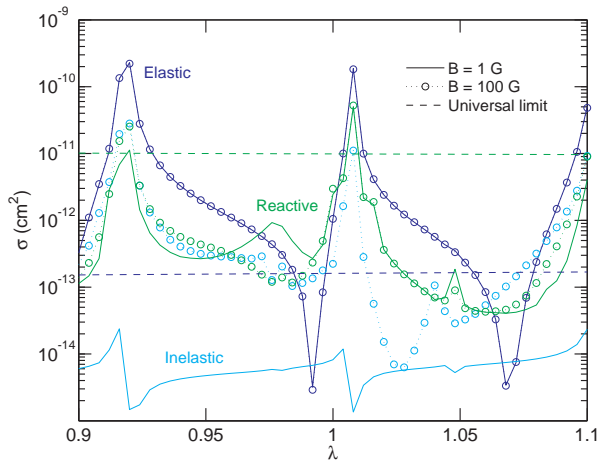}
\caption
  {\label{fig6:res4}
Cross sections for magnetically trapped $^{15}$NH + $^{15}$NH as a function of
the scaling factor $\lambda$, calculated for two magnetic field strengths
($B=1$ and 100 G) at $E = 10^{-6}$ K.  The results of the universal
quantum-defect model are also shown.
}
\end{figure}

An alternative way to establish the degree of universality is to test the
effect of small modifications in the (short-range) potentials. If the
scattering is universal, the cross sections are completely determined by the
long-range features of the interaction potentials. For instance, a scaling of
the potentials by a factor of $\lambda$ (or, equivalently, a scaling of the
reduced mass by $\lambda$) should change the universal elastic cross section by
$\lambda^{1/2}$.  Figure \ref{fig6:res4} shows the universal quantum-defect
results as a function of the scaling parameter $\lambda$ ($0.9 \leq \lambda
\leq 1.1$) for $E = 10^{-6}$ K.  The corresponding numerical cross sections,
obtained by reduced-mass scaling, are also shown for  $B=1$ and 100 G.  It is
evident that the numerical results are highly sensitive to the details of the
potentials, and that the universal model is inaccurate for all values of
$\lambda$.  In fact, the resonance features in the numerical cross sections are
signatures of non-universal behavior \cite{idziaszek:10a} and highlight the
importance of short-range physics in the dynamics. We thus conclude that the
scattering properties of magnetically trapped NH cannot be captured in a
universal model.

As discussed extensively in Refs.\ \cite{zuchowski:11,janssen:11,janssen:11a},
the $\lambda$-scaling approach also provides a means to sample the effects of
uncertainties in the interaction potentials. Although the potentials have been
obtained from state-of-the-art methods, there is a small remaining inaccuracy
in the \textit{ab initio} data which gives rise to an uncertainty in the
calculated cross sections.  This also carries implications for the prospects
for molecular evaporative cooling.  More specifically, one should evaluate the
elastic-to-reactive cross section ratio for all relevant $\lambda$ values to
obtain a realistic estimate of the cooling efficiency.  In the case of
$\lambda=1$, we find that reactive NH + NH collisions are more probable than
elastic ones (cf.\ Figs.\ \ref{fig6:res2} and \ref{fig6:res3}), while for
$\lambda\approx 0.95$ and $\lambda\approx1.03$ the elastic cross sections are
about one order of magnitude larger than the reactive ones.  These results
suggest that evaporative cooling of magnetically trapped NH might still be
feasible, but the probability of success is significantly smaller than
estimated earlier from non-reactive scattering calculations.  We note that
these findings are still valid when the size of the channel basis set is
increased, as detailed in the Supplementary Material.

It can also be seen in Fig.\ \ref{fig6:res4} that the inelastic cross sections
change rather dramatically from $B=1$ to 100 G, while the reactive cross
sections show only a weak dependence on magnetic field.  Nevertheless, for
certain values of $\lambda$, the reactivity can increase by almost one order of
magnitude as the magnetic field strength is changed. Thus, it may be possible
to control the NH + NH reaction rate by means of an external field. For most
$\lambda$ values, however, the reactive cross sections show $B^0$ behavior and
magnetic field control will not be possible. We note that the final
product-state distribution might be more sensitive to the magnetic field
strength than the \textit{total} reaction probability, but our scattering
method does not allow the calculation of product-state-resolved reaction cross
sections.

In summary, we have presented the first rigorous quantum scattering study of
ultracold reactive molecular collisions in the presence of a magnetic field.
Our results illustrate the importance of chemical reactions in a magnetically
trapped molecular gas, and call to reconsider the prospects for molecular
evaporative cooling and magnetically-controlled cold chemistry.  This work may
serve as a benchmark for other ultracold paramagnetic and dipolar molecules,
the dynamics of which are still virtually unexplored.

LMCJ and GCG thank the Council for Chemical Sciences of the Netherlands
Organization for Scientific Research (CW-NWO) for financial support. AvdA
thanks the Alexander von Humboldt Foundation for a Humboldt Research Award.

\end{document}